\documentclass[10pt]{article}
\usepackage[latin1]{inputenc}
\usepackage{amsmath}
\usepackage{amsfonts}
\usepackage{amssymb}
\usepackage{graphicx}
\usepackage{enumitem}
\usepackage{ctable}
\usepackage{natbib}
\usepackage[Lenny]{fncychap} 
\usepackage{fullpage}
\usepackage{float} 
\usepackage{caption}
\usepackage{subcaption}
\usepackage{makeidx}
\usepackage{xspace}
\usepackage[toc,page]{appendix}
\usepackage{boxedminipage}
\usepackage{hyperref}
\usepackage[all]{hypcap} % calls to \ref point to top of fig, not caption
\newcommand{\be}{\begin{enumerate}[noitemsep]}
\newcommand{\ee}{\end{enumerate}}
\begin{document}
%\setkeys{Gin}{width=0.95\linewidth} % was width = 0.95

\begin{center}
\Large
The Evolution of Lying in a Spatially-Explicit Prisoner's Dilemma Model

\normalsize
Gregg Hartvigsen

\href{mailto:hartvig@geneseo.edu}{hartvig@geneseo.edu}

Biology Department, SUNY Geneseo (Emeritus)
\end{center}

%%%%%%%%%%%%%%%%%%%%%%%%%%%%%%%%%%%%%%%%%%%%%%%%%%%%%%%%%%%%%%%%%%
\section{Abstract}

I present the results from a spatial model of the prisoner's dilemma, played on a toroidal lattice. Each individual has a default strategy of either cooperating ($C$) or defecting ($D$). Two strategies were tested, including ``tit-for-tat'' (TFT), in which individuals play their opponent's last play, or simply playing their default play. Each individual also has a probability of telling the truth ($0 \leq P_{truth} \leq 1$) about their last play. This parameter, which can evolve over time, allows individuals to be, for instance, a defector but present as a cooperator regarding their last play. This leads to interesting dynamics where mixed populations of defectors and cooperators with $P_{truth} \geq 0.75$ move toward populations of truth-telling cooperators. Likewise, mixed populations with $P_{truth} < 0.7$ become populations of lying defectors. Both such populations are stable because they each have higher average scores than populations with intermediate values of $P_{truth}$. Applications of this model are discussed with regards to both humans and animals.

%%%%%%%%%%%%%%%%%%%%%%%%%%%%%%%%%%%%%%%%%%%%%%%%%%%%%%%%%%%%%%%%%%
\section{Introduction}

There is an extensive body of research investigating the problem of how cooperation might evolve and persist in humans and other animals, where defecting, or cheating, should lead to higher overall returns \citep{Axelrod1981, Sigmund1993GamesOfLife,Dugatkin1997, Hartvigsen00, Lehmann2006}. For instance, lions (\textit{Panthera leo}) that cheat by not participating in a hunt reduce risk and yet benefit from successful kills conducted by members of their pride \citep{ScheelPacker1991}. Trivers (\citeyear{Trivers1971}) argued that cooperation could most easily evolve in systems where individuals recognized each other, allowing for reciprocity, or what is also considered to be ``contingent cooperation'' \citep[see][]{Cheney2011}. The model presented here introduces the effect of allowing truthfulness of prior behavior to evolve between interacting agents over time.

A very well studied model of this problem is the prisoner's dilemma (PD) in the field of game theory \citep[e.g.,][]{Rapoport1965}. The game is played between two individuals who have the choice of either cooperating ($C$) or defecting ($D$) \citep[see the broad exploration reviewed by][]{Poundstone1993}. The game was first described in 1950 by Flood and Drescher of the RAND Corporation and then later named by the Canadian mathematician Albert Tucker \citep{Poundstone1993,Kollock98}. Extensions of this game are vast with an effort, generally, to investigate the conditions by which cooperative behavior can be realized when defection is both a rational strategy and results in the highest individual payout. 

In the PD model, two cooperators will each receive a reward ($C_C$). A $C$ receives the lowest, sucker's payout ($C_D$) for playing against $D$ while the $D$ receives the highest, temptation payout $D_C$. Finally, two $D$ players each receive a punishment payout ($D_D$). This results in the Prisoner's dilemma where $D_C > C_C > D_D > C_D$. The rational choice in this game is to defect, regardless of the choice of their opponent, therefore resulting in a greater payoff. The dilemma arises because two cooperators will receive the highest average score since $\frac{C_C + C_C}{2} > \frac{C_D + D_C}{2}$ \citep{Rapoport1965, Axelrod1981}.

Interestingly, a tournament held by Axelrod led to the discovery of a winning strategy, entered by the mathematician Rapoport \citep{Axelrod1980EffectiveChoice}, which was called ``tit-for-tat.'' To play this strategy an individual begins by cooperating and then plays what the opponent played last. Although a very successful strategy, it has been shown to be defeated by other strategies, for example the win-stay, lose-shift model of Nowak and Sigmund (\citeyear{Nowak1993}).

Much of the work on this game comes from behavioral studies, including those with humans \citep{Kollock98} as well as animals (e.g., the evolution of sex ratios, Hamilton, \citeyear{Hamilton67}). Other examples for which this model could be applied include cooperation between humans and animals, such as the relationship between the Yao peoples of the Republic of Mozambique and the greater honeyguide (\textit{Indicator indicator}) bird \citep{Spottiswoode2016}. In this cooperative relationship both bird and humans are able to initiate the hunt through calls which result in the bird leading a hunter to a nest and the hunter then opening the nest to expose the honey and beeswax. Additionally, a variety of bird species exhibit warning calls that can reduce predation risk for nearby individuals, resulting in cooperation being favored among relatives, referred to as kin selection \citep{Hamilton1964a}. Another example for which cooperation leads to high payoffs and defection to low payoffs for both participants includes the system of cleaner fish that remove parasites from hosts \citep{Poulin1996, Gingins2013}. 

The prisoner's dilemma game also has been investigated in spatially-explicit environments, such as on a two-dimensional lattice, as is done in this work. These models require the model to be run repeatedly, or iterated, so that individuals can implement the tit-for-tat strategy. These models can be rich with dynamical outcomes, depending on the relative payouts for cooperators and defectors, including the possibility of creating persistent cycles of cooperators and defectors \citep{Nowak-May1992}.

One consideration, not yet investigated, involves whether a player's last play is honestly conveyed to an opponent. In this paper I investigate how truth-telling, and the associated ability to lie, might be used to increase individual scores. The games played here allow each player to play twice against each of the four neighbors for a total of eight plays each. Each cell is then replaced by a neighbor whose probability of telling the truth ($P_{truth}$) can be modified through mutation and, ultimately, lead to evolution of truth-telling over time. In this exploration, individuals either play their default strategy (they're either a cooperator or defector), or they play tit-for-tat. The goal of this paper, therefore, is to assess the conditions that lead to populations moving either toward or away from truth-telling in order to maximize their individual scores and, ultimately, whether communities achieve evolutionary stable strategies of global cooperation or defection \citep[sensu,][]{MaynardSmith1982EvolutionaryGames}.

%%%%%%%%%%%%%%%%%%%%%%%%%%%%%%%%%%%%%%%%%%%%%%%%%%%%%%%%%%
\section{Methods}

A spatial prisoner's dilemma model was developed using a 40x40 toroidal lattice. Each cell of the lattice represents an individual player (target) with four opponents. The opponents may be nearest neighbors (a von Neumann neighborhood) or randomly chosen from anywhere in the lattice (a non-spatial approximation). Therefore, each individual has four opponents and will serve as one of the opponents for each of those four individuals.

\begin{figure}[H]
\begin{center}
\includegraphics[width = 4cm]{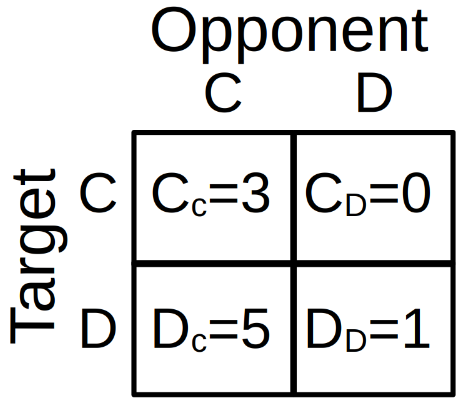}
\caption{Payoffs to the target in the prisoner's dilemma game used in this paper. Letters represent cooperators (C) and defectors (D) with the subscript representing the opponent. Note that regardless of what the opponent plays, the target receives a higher payout by defecting.}
\label{payout-matrix}
\end{center}
\end{figure}

During a round, each player is chosen as a target individual to play against their four opponents. Each play is scored using a standard prisoner's dilemma payout matrix for both players (see figure \ref{payout-matrix}). After all individuals have served as a target individual the simulation advances to the reproduction phase (see Reproduction section \ref{reproduction}). 

%vvvvvvvvvvvvvvvvvvvvvvvvvvvvvvvvvvvvvvvvvvvvvvvvvvvvv
%^^^^^^^^^^^^^^^^^^^^^^^^^^^^^^^^^^^^^^^^^^^^^^^^^^

Simulations start with either all defectors, all cooperators, or a random setup where individuals are chosen with equal probability to be either a cooperator or defector. Individuals are assigned the same initial probabilities of telling the truth ($IPT$, see values tested in Table 1). The model tests two strategies for play, including tit-for-tat and simply play an individual's default strategy. For the tit-for-tat strategy, individuals play what their neighbors present as their last play, which would be the opposite play if a uniformly distributed random number is less than or equal to their $P_{truth}$. If the neighbor has not played previously they present their default play unless, again, they lie.

%vvvvvvvvvvvvvvvvvvvvvvvvvvvvvvvvvvvvvvvvvvvv
\begin{table}[H]
\begin{minipage}[t]{1\linewidth}
\caption{Parameters and settings tested in the main experiment (see Table 2). The mutation rate affects only $P_{truth}$, allowing it to evolve over time.}
\begin{center}
\begin{tabular}{|l|l|}
\hline
Parameter & Settings tested \\
\hline
\hline
Initial setup & all C, all D, random C and D\\
\hline
Initial $P_{truth}$ ($IPT$) & sequence from 0 $\rightarrow$ 1 by 0.025\\
\hline
Strategies & Tit-for-tat, play default\\
\hline
Mutation rate & 0 or 0.1\\
\hline
\end{tabular}
\label{parm-settings}
\end{center}
\end{minipage}
\end{table}
%^^^^^^^^^^^^^^^^^^^^^^^^^^^^^^^^^^^^^^^^^^^^

Additional simulations were completed to investigate invasibility and where values for the initial probabilities of truth-telling ($IPT$) were chosen randomly. The invasibility simulations started with a single, unique individual that differed from the population in strategy and/or $P_{truth}$. This leads to 16 possible combinations of invader and population. However, no invasibility simulations were run where the invader matched the population for strategy and $P_{truth}$. Additionally, for visualization purposes, a 50 x 50 lattice was divided in half with cooperators on one half and defectors on the other, with simulations showing what spread looks like under different conditions. All combinations of factors were replicated three times and simulations were run for 1000 time steps. The model and statistical analyses were completed using the R Language and Environment for Statistical Computing \citep{R}.

%----------------------------------------------------
\subsection{The algorithm for playing a single round}

During the beginning of a round all scores are zeroed. Each individual is chosen systematically and play by the following rules. Note that each individual plays their four neighbors as the target individual and then serves as an opponent for their four neighbors. Each individual is assigned to be either a $C$ or $D$, which is used during the default strategy simulations or when an individual has not previously played.

\be
\item Choose a target individual and collect their four neighbors on the lattice.
\be
\item For the tit-for-tat strategy simulations: 
\be
\item the target individual plays what each opponent presents as their last play, which might be incorrect if a uniformly distributed random number is less than or equal to each opponent's $P_{truth}$. 
\item If an opponent has not played previously then they use their default play which, again, may or may not be truthfully presented because of their $P_{truth}$.
\ee
\item For the default strategy simulations individuals simply use their default play and do not query opponents for their previous play so $P_{truth}$ is not used.
\ee
\item After each interaction the individual's play is stored and the payoff obtained recorded (figure \ref{payout-matrix}).
\ee

\noindent Once all individuals have served as a target they will each have played eight times. The scores for each individual are then used to determine their ability to reproduce in the next time step. 

%----------------------------------------------------
\subsection{Reproduction}\label{reproduction}

After a round is completed each cell is replaced by one of its four neighbors, based on their scores using a random, roulette wheel algorithm which preferentially chooses a neighbor with the highest score. The new, replacement individuals for each cell are stored separately from the original lattice so that updating of the population is discrete. This replacement is then done at the end of each round.

If the mutation rate is set to 0.1 a uniformly distributed random number is selected and, if it is less than or equal to that mutation rate, then the individual's $P_{truth}$ is either increased or decreased, with equal probability, by 0.05. The truth-telling probability is bounded ($0 \leq P_{truth} \leq 1.0$). If the mutation rate parameter is set to zero no changes are made to $P_{truth}$ (the condition of ``no evolution'').

%%%%%%%%%%%%%%%%%%%%%%%%%%%%%%%%%%%%%%%%%%%%%%%%%%%%%%
\section{Results}

\begin{figure}[H]
\begin{center}
\includegraphics[width = \textwidth]{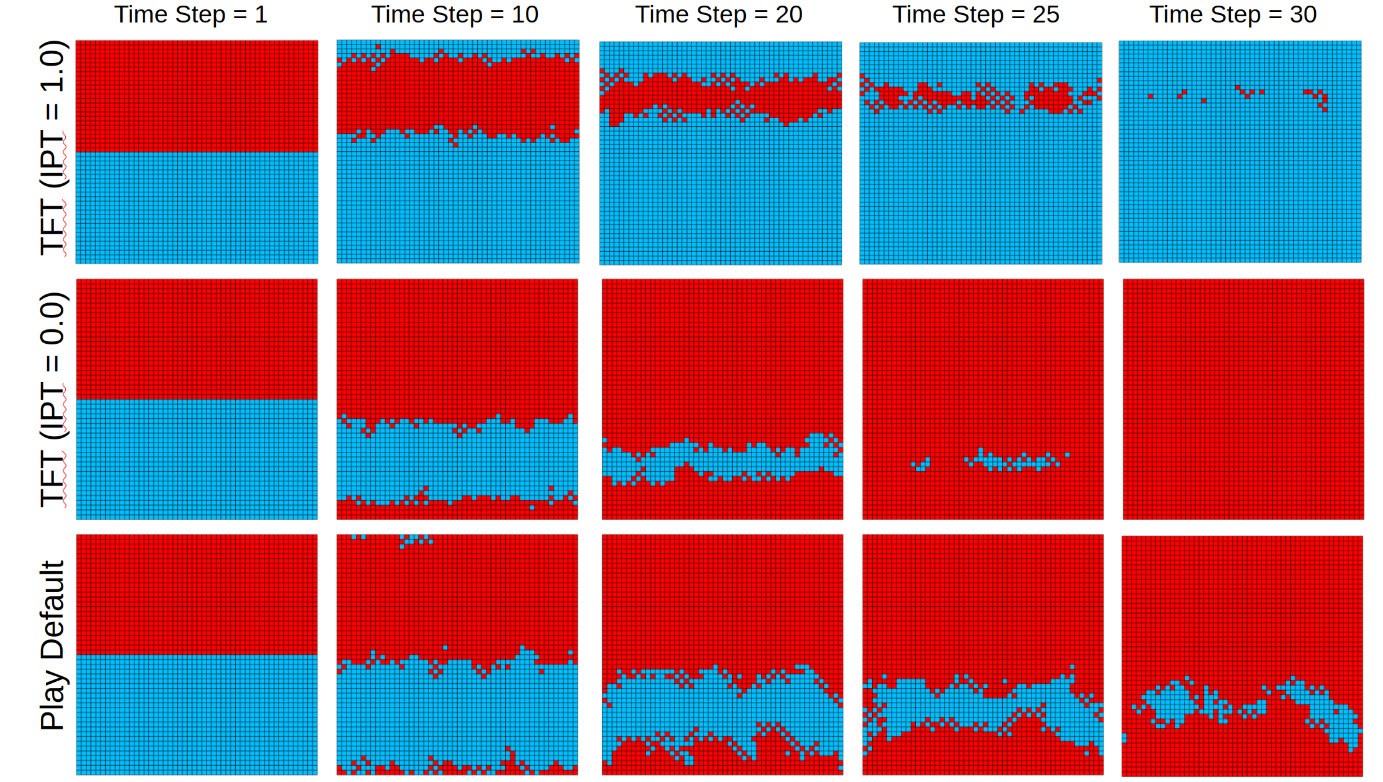}
\caption{Sample simulations over time for three different conditions, including playing tit-for-tat (TFT) with an initial probability of truth-telling ($IPT$) of 1.0 (row 1), TFT with $IPT = 0.0$ (row 2), and individuals playing their default strategies (row 3). Note that cooperators are blue (initially lower half of lattice) and defectors are red. For the TFT simulations, $P_{truth}$ was allowed to evolve. For the default strategy there is no querying of opponents so $P_{truth}$ does not play a role in those interactions. For this visualization, simulations were run on 50 x 50 toroidal lattice.}
\label{grid.fig}
\end{center}
\end{figure}

Regardless of initial conditions or whether the truth-telling parameter ($P_{truth}$) evolves over time, communities reach the state of either all cooperators or all defectors. For instance, a community with half cooperators and half defectors that all play tit-for-tat and initially tell the truth ($IPT$ = 1.0) will end with truth-telling cooperators ($P_{truth} \approx 1.0$, row 1, figure \ref{grid.fig}). In contrast, the same community that starts with a mixed population but with $IPT$ = 0.0 will end with all lying defectors ($P_{truth} \approx 0.0$, row 2, figure \ref{grid.fig}). If individuals do not query their neighbors and simply play their default strategy ($IPT$ is irrelevant), the defectors take over the community (row 3, figure \ref{grid.fig}).  

%vvvvvvvvvvvvvvvvvvvvvvvvvvvvvvvvvvvvvvvvvvvvvvvvvvvvvvvvvv
%^^^^^^^^^^^^^^^^^^^^^^^^^^^^^^^^^^^^^^^^^^^^^^^^^^^^^^^^^

Five factors were tested for their influence on average final scores and $P_{truth}$ after 1000 time steps. All factors and their interactions were statistically significant. The least important factor, however, was the test of the importance of spatial dependence, which tested the difference in model dynamics when using nearest neighbor opponents versus randomly chosen neighbors for interactions. Given this, the remaining analyses omit the non-spatial simulations.

The proportions of variance explained by each factor, and their interactions, for final average score and $P_{truth}$, are shown in Table 2. Three of the effects for each of these response variables explained a relatively large proportion of the variance (highlighted and with an asterisk in Table 2). These are further discussed below.

%vvvvvvvvvvvvvvvvvvvvvvvvvvvvvvvvvvvvvvvvvvvvvvvvvvvvvvvvvvvvvvvvvvvvvv
\begin{table}[H]
	\begin{center}
		\caption{Analysis of variance table showing the degrees of freedom and the proportion of the variance ($\frac{SS_i}{\sum{SS}}$) explained by each $i$ factor, and their interactions, for the mean score and mean $P_{truth}$ response variables. The six relatively high proportions that are highlighted and starred are further discussed in the text. See Table 1 for the the levels set for each factor. $IPT$ stands for initial values of $P_{truth}$. Note that all terms were significant at the p $<$ 0.001 level.}
		\begin{tabular}{|l|c|c|c|}
			\hline
			Factor                &   Df & Score & $P_{truth}$ \\
			\hline
			\hline
			Setup (allC, allD, random)        &  2 & \textbf{0.516*} & 0.088 \\ 
			\hline
			$IPT$ (0 $\rightarrow$ 1, by 0.025)       &    40 & 0.003 & \textbf{0.367*} \\ 
			\hline
			Strat (play default or play TFT) &     1 & \textbf{0.226*} & 0.003 \\ 
			\hline
			Evol (yes or no)                  &     1  &  0.016 & 0.003 \\ 
			\hline
			Setup:$IPT$             &    80   &  0.017 & 0.023 \\ 
			\hline
			Setup:Strat          &      2  &  \textbf{0.158*} & \textbf{0.102*} \\ 
			\hline
			$IPT$:Strat             &    40  &  0.003 & 0.006 \\ 
			\hline
			Setup:Evol            &     2   &  0.004 & 0.088 \\ 
			\hline
			$IPT$:Evol              &    40   &  0.002 & 0.089 \\ 
			\hline
			Strat:Evol            &     1   &  0.016 & 0.003 \\ 
			\hline
			Setup:$IPT$:Strat       &   80   & 0.017 & 0.029 \\ 
			\hline
			Setup:$IPT$:Evol        &   80   & 0.008 & 0.023 \\ 
			\hline
			Setup:Strat:Evol      &    2   &  0.004 & \textbf{0.102*} \\ 
			\hline
			$IPT$:Strat:Evol        &   40   & 0.002  & 0.006\\ 
			\hline
			Setup:$IPT$:Strat:Evol  &   80   & 0.008  & 0.029\\ 
			\hline
		\end{tabular}
	\end{center}
\end{table}
%^^^^^^^^^^^^^^^^^^^^^^^^^^^^^^^^^^^^^^^^^^^^^^^^^^^^^^^^^^^^^^^^^

\begin{figure}[H]
\begin{center}
\includegraphics{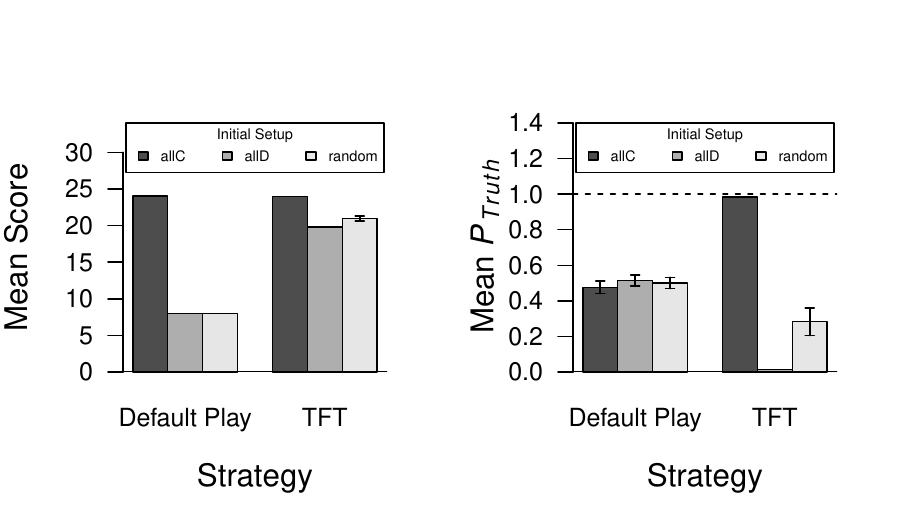}
\caption{The mean score (left) and mean $P_{truth}$ (right) at the end of simulations that ran for 1000 time steps on a 40 x 40 lattice. Only simulations with mutation = 0.1 (evolution) are included. For the default strategy scores (left graph), the communities end as either all cooperators or all defectors, leading individuals to having scores of either 24 or 8, respectively. When relying on their default strategy, the $P_{truth}$ follows a random walk (right graph, and see figure \ref{randTP-time.fig}). Note that in the mean $P_{truth}$ graph (right panel) the random initial setup, with individuals playing TFT, had an average less than 0.5 because runs that start with $IPT$ values less than 0.7 decay toward zero (see lower-left panel of figure \ref{Truth.TFT}). Error bars are 95\% confidence intervals.}
\label{setup-by-strat.fig}
\end{center}
\end{figure}

\begin{figure}
\begin{center}
\includegraphics{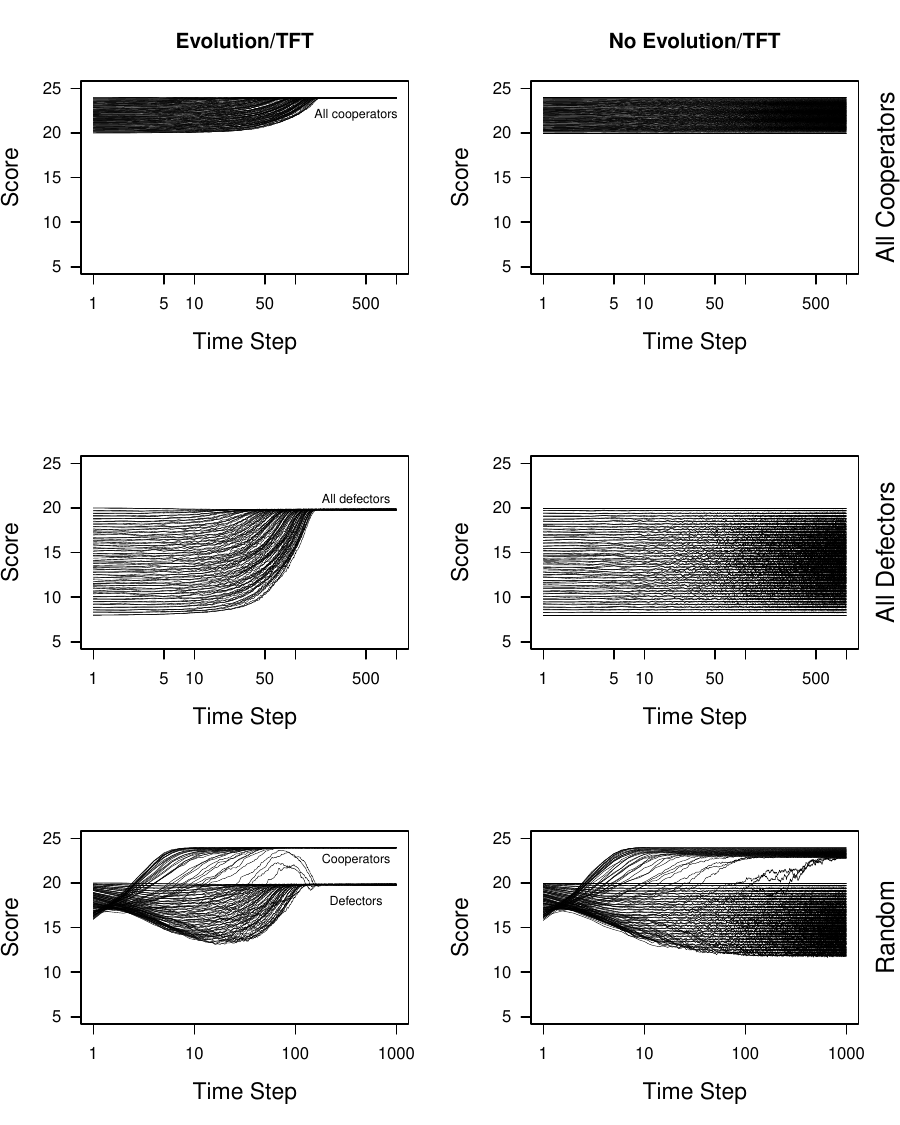}
\caption{Mean scores over time for simulations where $P_{truth}$ does (left panels) and does not (right panels) evolve. All individuals play the tit-for-tat strategy and start with all cooperators (row 1), all defectors (row 2), and a random distribution of cooperators and defectors (row 3). Note that the x-axis scale is logged.}
\label{Score.TFT}
\end{center}
\end{figure}

\begin{figure}
\begin{center}
\includegraphics{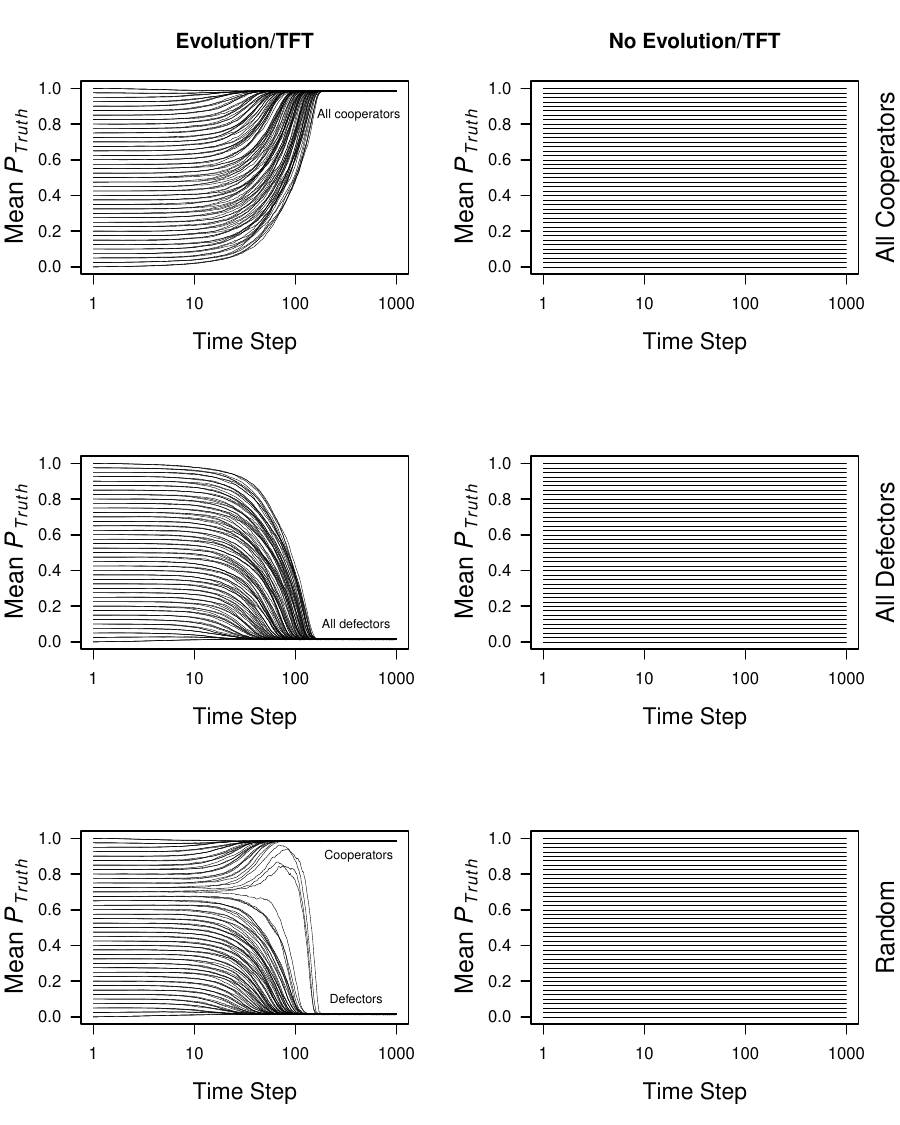}
\caption{The mean $P_{truth}$ over time for simulations for which $P_{truth}$ does (left panels) and does not (right panels) evolve. Initial $P_{truth}$ ($IPT$) values range from 0 to 1, by 0.025. All individuals use the tit-for-tat strategy and start with all cooperators (row 1), all defectors (row 2), and a random distribution of cooperators and defectors (row 3). $P_{truth}$ values in the left panels do not end at either 0.0 or 1.0 exactly due to mutation. Note that the x-axis scales are logged.}
\label{Truth.TFT}
\end{center}
\end{figure}

\begin{figure}
\begin{center}
\includegraphics{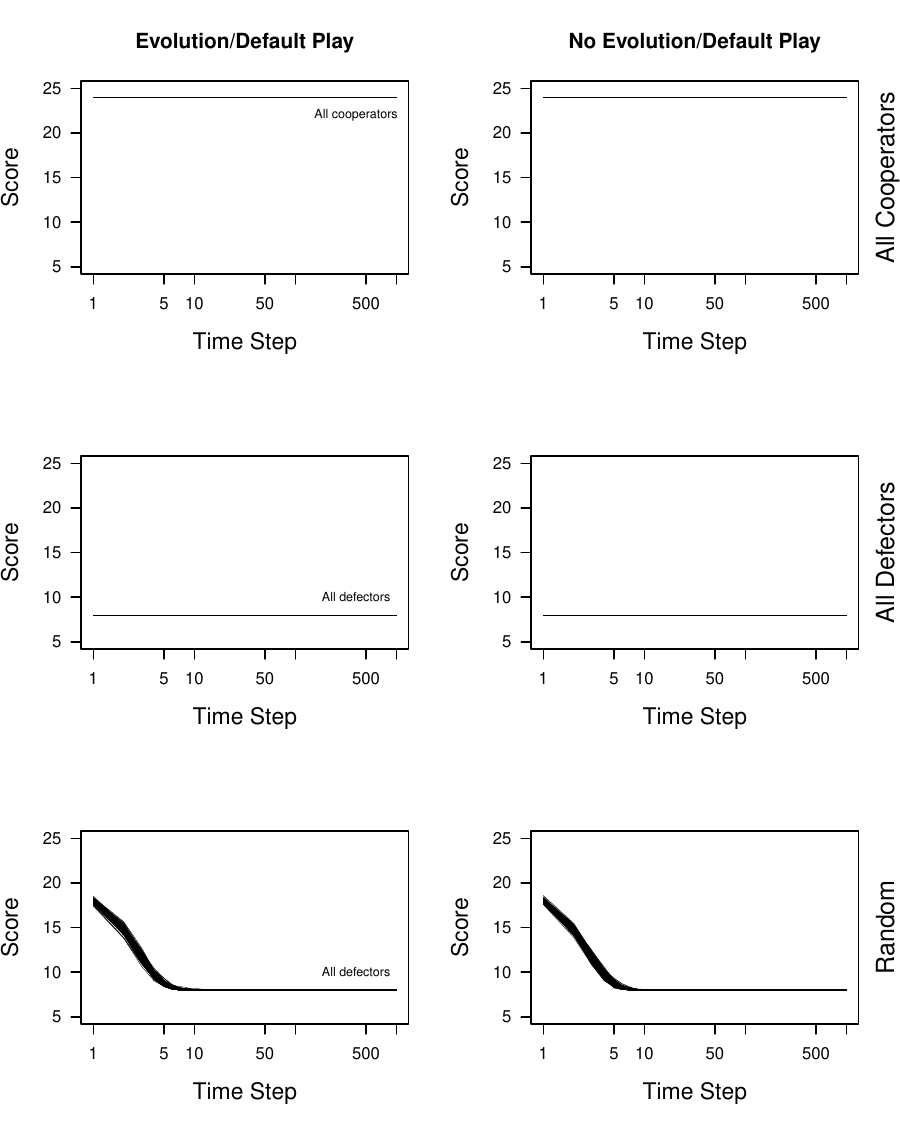}
\caption{Mean average scores over time for simulations for which $P_{truth}$ does (left panels) and does not (right panels) evolve. Individuals play their default strategies and start with all cooperators (row 1), all defectors (row 2), and a random distribution of cooperators and defectors (row 3). Note that the x-axis scale is logged.}
\label{score-default.fig}
\end{center}
\end{figure}

\begin{figure}
\begin{center}
\includegraphics{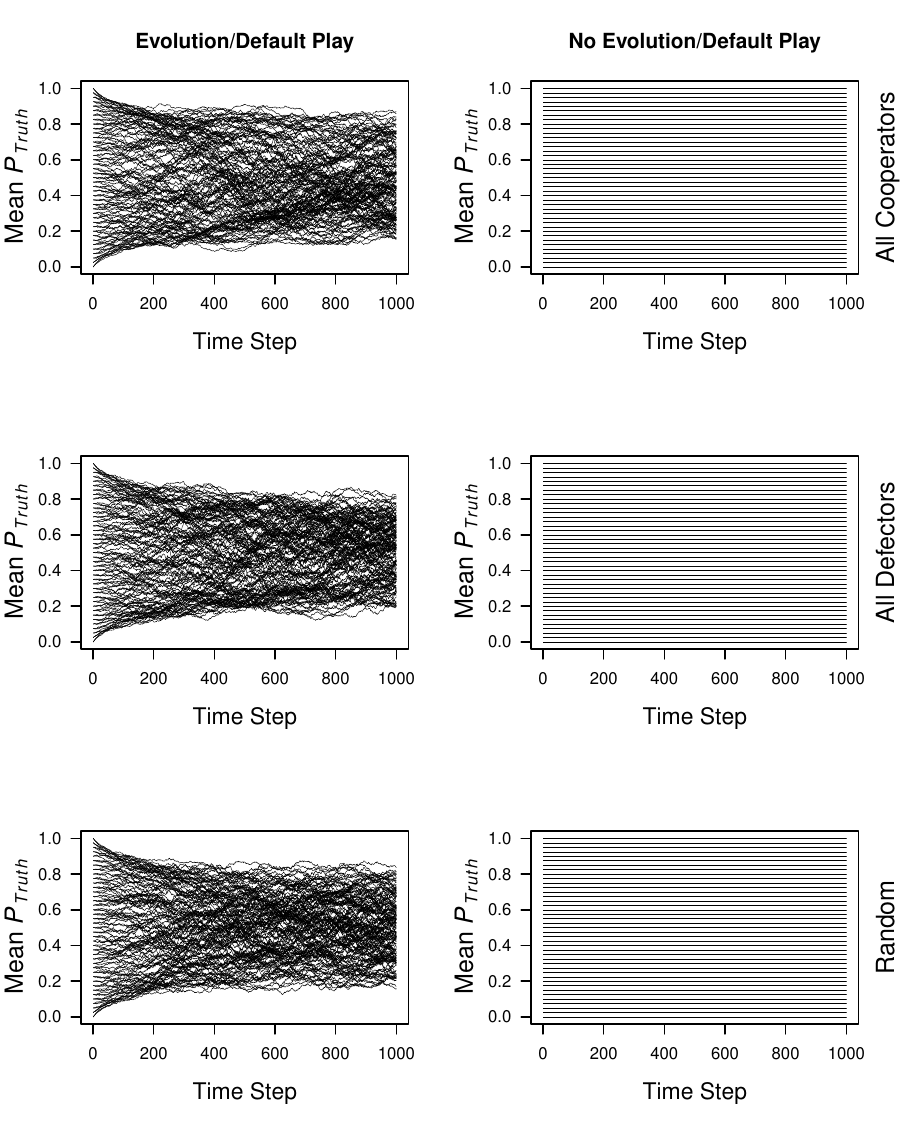}
\caption{The mean $P_{truth}$ over time for simulations for which $P_{truth}$ does (left panels) and does not (right panels) evolve. In these simulations individuals play their default strategy, regardless of their probability of telling the truth. For all three initial conditions (allC, allD, and random mix of cooperators and defectors) there is no change in the truth probabilities over time without evolution. However, with evolution the truth probability is not used by individuals so the changes represent simple random walk dynamics. Note that the x-axis scale is not logged.}
\label{truth.play.default}
\end{center}
\end{figure}

The greatest effect on the final score was due to the initial setup of the communities (see left graph, figure \ref{setup-by-strat.fig}). This was because cooperating communities would reach average scores of 24 (all truth-telling). However, simulations with individuals using only their default strategy (left side of left panel, figure \ref{setup-by-strat.fig}) would move toward pure defection where defectors all earn a score of one during each of their eight interactions (figure \ref{payout-matrix}). For TFT simulations lying defectors were able to increase their scores because they would present as a cooperator and their opponent would then cooperate, yielding a higher score, seen in the right bars in the left panel of figure \ref{setup-by-strat.fig}. However, they still did not score as high as the truth-telling cooperators. The dynamics of these scenarios over time are shown in figures \ref{Score.TFT}--\ref{truth.play.default}

The interaction between setup and strategy also was important for the final average truth probability (Table 2 and right graph, figure \ref{setup-by-strat.fig}). For the truth probability with the default play, the values actually behave as a random walk with evolution because no selection is operating (individuals are not querying each other for their last play). This also can be seen in the left panels of figure \ref{truth.play.default}. When playing TFT the cooperators end with high truth-telling values while the defectors end with values near zero (not exactly zero because of mutation). Finally, the simulations that begin with randomly placed cooperators and defectors end with average truth-telling values below 0.5 because simulations that have initial truth-telling values below 0.75 end as lying communities (right-most bar in right graph of figure \ref{setup-by-strat.fig}).

%vvvvvvvvvvvvvvvvvvvvvvvvvvvvvvvvvvvvvvvvvvvvvvvvvvvvvvvvvv
%\setkeys{Gin}{width=1\linewidth} % was width = 0.95
%\setkeys{Gin}{width=0.9\linewidth} % was width = 0.95
%^^^^^^^^^^^^^^^^^^^^^^^^^^^^^^^^^^^^^^^^^^^^^^^^^^^^^^^^^

%\setkeys{Gin}{width=0.9\linewidth} % was width = 0.95
%vvvvvvvvvvvvvvvvvvvvvvvvvvvvvvvvvvvvvvvvvvvvvvvvvvvvvvvvvvvvvvvvvvvv
% TFT Score
%^^^^^^^^^^^^^^^^^^^^^^^^^^^^^^^^^^^^^^^^^^^^^^^^^^^^^^^^^^^^^^^^^

%vvvvvvvvvvvvvvvvvvvvvvvvvvvvvvvvvvvvvvvvvvvvvvvvvvvvvvvvvvvvvvvvvvvv
% TFT Truth Probability

%^^^^^^^^^^^^^^^^^^^^^^^^^^^^^^^^^^^^^^^^^^^^^^^^^^^^^^^^^^^^^^^^^

%vvvvvvvvvvvvvvvvvvvvvvvvvvvvvvvvvvvvvvvvvvvvvvvvvvvvvvvvvvvvvvvvvvvv
%^^^^^^^^^^^^^^^^^^^^^^^^^^^^^^^^^^^^^^^^^^^^^^^^^^^^^^^^^^^^^^^^^

%vvvvvvvvvvvvvvvvvvvvvvvvvvvvvvvvvvvvvvvvvvvvvvvvvvvvvvvvvvvvvvvvvvvv
%^^^^^^^^^^^^^^^^^^^^^^^^^^^^^^^^^^^^^^^^^^^^^^^^^^^^^^^^^^^^^^^^^
%\setkeys{Gin}{width=0.95\linewidth}

Additionally, there was a relatively strong three-way interaction of setup, strategy, and evolution. Evolution by itself was a highly significant factor although it explained little in the overall scores or truth-telling values (Table 2). The reason for this is that the analysis of variance only takes into account the final values. If, however, we look at the time series data the effect of evolution on scores and $P_{truth}$ we can see it's effect in allowing these response variables to change (figures \ref{Score.TFT}--\ref{truth.play.default}).

%--------------------------------------
\subsection{What happens to $P_{truth}$ when $IPT$ values are randomly assigned?}

\begin{figure}
\begin{center}
\includegraphics{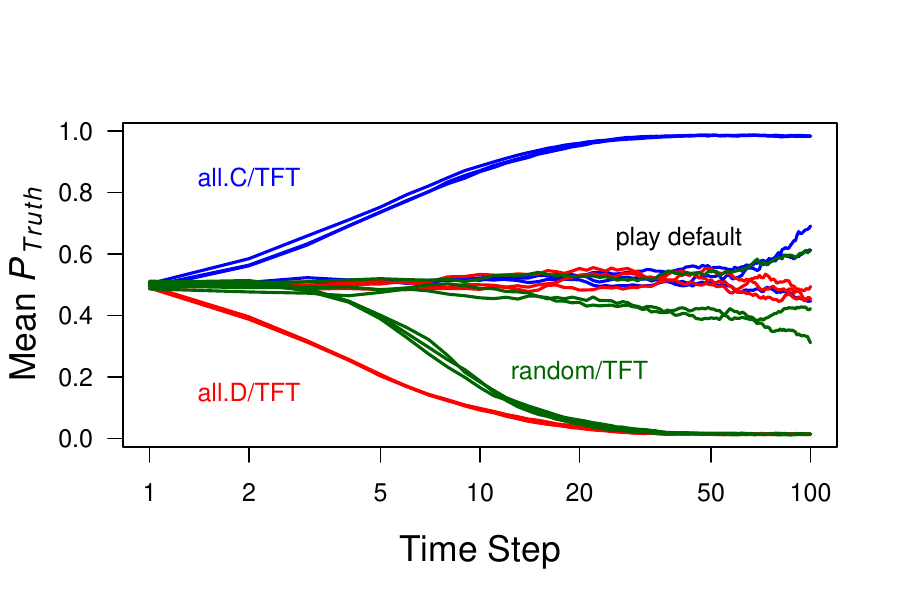}
\caption{The average $P_{truth}$ over time for simulations beginning with uniformly distributed random values and, therefore, a mean of 0.5. The average $P_{truth}$ increases in communities with all cooperators playing tit-for-tat (all.C/TFT) and decreases with all defectors (all.D/TFT). Interestingly, when cooperators and defectors are randomly mixed in the community (``random/TFT''), the population evolves toward all liars with the defectors filling the community. As expected, when individuals play their default strategies the mean $P_{truth}$ follows a random walk, diverging slowly from 0.5 for all initial setups. Each test was replicated three times on a 40 x 40 lattice.}
\label{randTP-time.fig}
\end{center}
\end{figure}

When $P_{truth}$ is initially assigned to individuals using a uniform random distribution the values follow a typical random walk when individuals simply play their default strategies (figure \ref{randTP-time.fig}). This is because individuals do not query their neighbors for their last play. However, when playing TFT and the mutation rate of $P_{truth}$ is 0.1, allowing for evolution, the mean $P_{truth}$ evolves toward 1.0 in cooperative communities and toward 0.0 in defector communities (figure \ref{randTP-time.fig}). If individuals are initially randomly assigned as cooperators and defectors, truth-telling erodes and lying defectors take over communities.

%\setkeys{Gin}{width=0.75\linewidth}
%vvvvvvvvvvvvvvvvvvvvvvvvvvvvvvvvvvvvvvvvvvvvvvvvvvvvvvvvvvvvvvvvvvvvvvvvvvvvvvvv
%^^^^^^^^^^^^^^^^^^^^^^^^^^^^^^^^^^^^^^^^^^^^^^^^^^^^^^^^^^^^^^^^^^^^^^^^^^^^
%\setkeys{Gin}{width=0.95\linewidth}

%--------------------------------------
\subsection{Cooperators tell the truth and defectors lie}

\begin{figure}
\begin{center}
\includegraphics{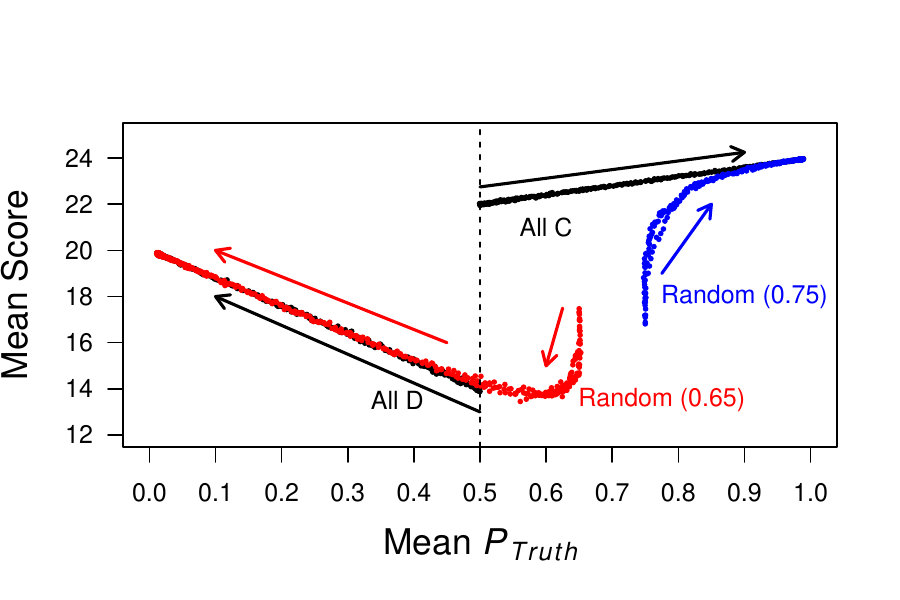}
\caption{The phase-plane relationship between mean score and the mean $P_{truth}$ over time (the arrows indicate direction of change). The four different scenarios are each represented by three independent simulations, with each running for 1000 time steps. The ``All C'' and ``All D'' simulations began with $P_{truth} = 0.5$ for each individual. The two random scenarios began with $P_{truth}$ set as shown. Individuals in all four scenarios played tit-for-tat with the probability of truth-telling able to evolve over time. Note that all simulations eventually move toward increasing scores despite the mean $P_{truth}$. For random setups there is a threshold that lies between 0.65 and 0.75.}
\label{score-vs-truth}
\end{center}
\end{figure}

In simulations where individuals play TFT and $P_{truth}$ evolves we find that communities of defectors decline raapidly toward $P_{truth} = 0$ (figure \ref{score-vs-truth}). In cooperator communities $P_{truth}$ increases toward 1.0. These two sets of simulations, labeled ``All D'' and ``All C'' in figure \ref{score-vs-truth}, both exhibit increasing average scores. Therefore, these conditions reach stable equilibria where any further change in $P_{truth}$ results in decreased scores.

In simulations with a random mixture of cooperators and defectors playing TFT, the model reveals a threshold between lying defector or truth-telling cooperator communities (figure \ref{score-vs-truth}). For populations that have an initial $P_{truth} = 0.65$, the populations move toward all defectors with $P_{truth} \rightarrow 0.0$. Likewise, if similar populations begin with an initial $P_{truth} = 0.75$ the individuals will move toward all cooperators that end with $P_{truth} \rightarrow 1.0$ (figure \ref{score-vs-truth}). Each of these sets of populations achieve stable equilibria.

Interestingly, in figure \ref{score-vs-truth} the scores for the  ``Random (0.65)'' populations initially decline. This seems counter-intuitive since selection should favor populations that increase their scores. However, these populations begin with an even mix of cooperators and defectors. The defectors, however, mostly tell the truth so cooperators defect, gaining a lower score. The defectors begin to spread and average scores decline. With decreasing values of $P_{truth}$ defectors begin to have higher scores with increased lying because these defectors present as cooperators but behave as defectors. As $P_{truth}$ continues toward zero the mean scores increase to the point that a lying defector can coerce their four neighbors to cooperate (+5 pts each) but then get the $C_D$ (sucker's) payout (+0) when they cooperate against their four lying defector neighbors (see figure \ref{payout-matrix}). Therefore, lyind defector populations end with mean scores of 20 (as opposed to 24 in truth-telling cooperator communities). The observed threshold for simulations going toward either all cooperators or all defectors is at $P_{truth} = 0.75$ because presenting play truthfully to a majority of neighbors allows cooperators to gain the highest scores and, therefore, spread in the lattice (figure \ref{score-vs-truth}).

%vvvvvvvvvvvvvvvvvvvvvvvvvvvvvvvvvvvvvvvvvvvvvvvvvvvvvvvvvvvvvvvvvvvv
%^^^^^^^^^^^^^^^^^^^^^^^^^^^^^^^^^^^^^^^^^^^^^^^^^^^^^^^^^^^^^^^^^

%----------------------------------------
\subsection{Community invasibility}

For a unique individual invader ($I$) to successfully invade a uniform population of a different type the invader needs to be able to replace a single neighbor $N$ with a probability $P(N_I) \geq 0.5$. We calculate this probability $P(N_I)$ that a neighbor of an invading individual is replaced by the invader using the ratio of the score the invader gets ($S_I$) playing as target and as an opponent against its four neighbors. We also calculate the score of the neighbor playing against its three like types and the invader ($S_N$ below). Here, this is considered only for communities where individuals are playing TFT and, therefore, querying their neighbors for their last play. After one round of play we can calculated the probability that the invader replaces one opponent. This follows the roulette wheel algorithm for how each cell is replaced each time step (see Methods).

\begin{equation}\label{invasibility.eq}
\begin{split}
S_I =&~ 4 \cdot I_{N_T} + 4 \cdot I_{N_O}\\
S_N =&~ N_{I_T} + N_{I_O} + 3 \cdot N_{N_T} + 3 \cdot N_{N_O}\\
P(N_I) =&~ \frac{S_I}{S_I + S_N}
\end{split}
\end{equation}

\noindent $S_I$ is the score acquired by the invader ($I$) playing its four neighbors ($N$) as target ($T$) and as opponent ($O$). $S_N$ is the score a neighbor of the invader gets by playing the invader and its three like-type neighbors as target ($N_{I_T}$) and as opponent ($N_{I_O}$). Additionally, the neighbor's score is affected by playing its remaining three like neighbors as target ($N_{N_T}$) and as opponent ($N_{N_O}$). The ratio ($P(N_I)$) is the probability a neighbor gets replaced by the invader. 

\begin{figure}
\begin{center}
\includegraphics[width = 6cm]{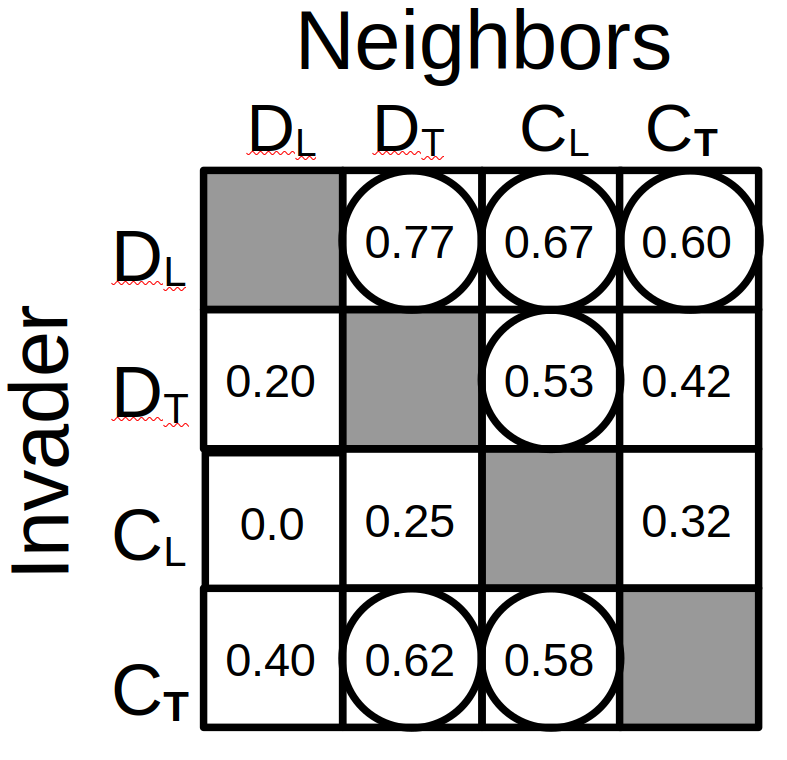}
\caption{The probabilities ($P(N_I)$) of an invading individual replacing a single neighbor (equation \ref{invasibility.eq}). Circled probabilities are conditions where invasion is possible ($P(N_I) \geq 0.5$). For instance, a lying cooperator ($C_L$) is not able to invade any community. This is because, in a tit-for-tat game, they will lie and present that they will defect causing their opponent to defect, and they ultimately cooperate, earning the score of 0 ($C_D$, figure \ref{payout-matrix}). Likewise, a defector community of pure liars ($D_L$) is uninvadable. Gray boxes are not considered since individuals do not invade their own communities.}
\label{invaders.fig}
\end{center}
\end{figure}

These probabilities for each type of interaction between invader and a population are shown in figure \ref{invaders.fig}. Interestingly, half of the interactions allow for a strategy to invade a population. Not surprisingly, however, there is a zero probability of a lying cooperator to invade a population of defecting liars. The highest probability for a successful invasion is for a lying defector ($D_L$) to invade a population of truth-telling defectors ($D_T$). In this scenario a lying defector presents as a cooperator so that the neighbor cooperates, and then the target defects, earning the highest payout (5 pts against each for four neighbors). Notably, the model suggests that a lying defector can invade any community. Interestingly, the scenario of a single $D_L$ invading a population of $C_T$ is portrayed in the movie ``The Invention of Lying,'' (2009). This model suggests that the $D_L$ strategy, playing TFT, will take over the community.

\begin{figure}
\begin{center}
\includegraphics{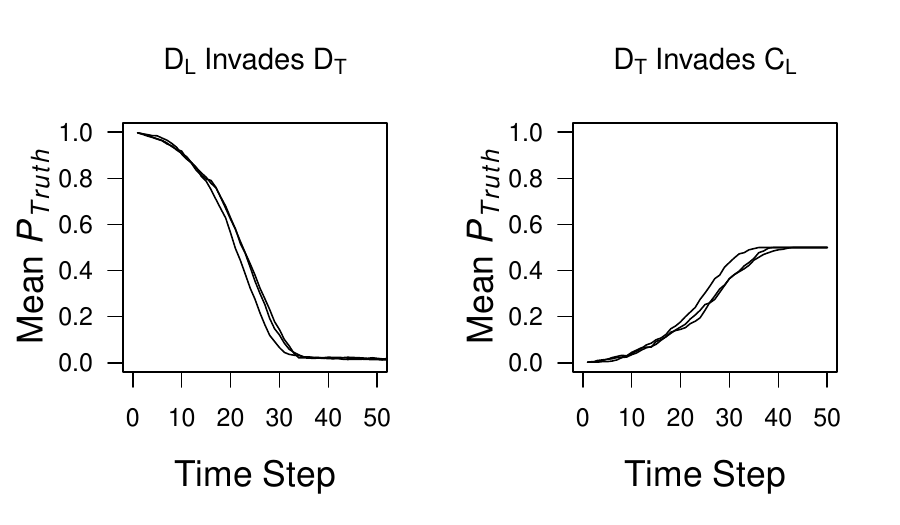}
\caption{In the left graph the mean $P_{truth}$ for three simulations decreased with a single lying defector ($D_L$) introduced into a population of truth-telling defectors ($D_T$). The right shows three simulations of the invasion of a single truth-telling defector ($D_T$) in a population of lying cooperators ($C_L$).}
\label{invader-test.fig}
\end{center}
\end{figure}

We can test these invasibility probabilities using the model, recognizing that a probability that is just over 0.5 may still lead to outcomes where an invasion fails. In figure \ref{invader-test.fig} we see that a lying defector ($D_L$) can lead to a community transition when invading a community of truth-telling defectors ($D_T$). The average truth-telling parameter then decreases so all individuals become lying defectors (left graph, figure \ref{invader-test.fig}). Additionally, a truth-telling defector can invade a community of lying cooperators (right graph, figure \ref{invader-test.fig}).

%%%%%%%%%%%%%%%%%%%%%%%%%%%%%%%%%%%%%%%%%%%%%%%%%%%%%%%%%%%%%%%%%%%%%%
\section{Discussion}

The prisoner's dilemma model is a very well studied game-theoretic model and is one of many types of social dilemma models \citep{Poundstone1993, Sigmund1993GamesOfLife, Kollock98, Lehmann2006} . Generally, it is a game played between a pair of individuals and, interestingly, the rational behavior of the players ultimately leads to a relatively poor outcome for both. This has led researchers to seek out conditions that might lead to cooperative behavior, which is observed in a variety of animal systems, including interactions among humans.

In the prisoner's dilemma there is a variety of strategies that individuals can use, including simply play fixed strategies, such as always cooperate or always defect. Alternatively, individuals can use mixed strategies. The now well-known strategy of tit-for-tat \citep{Axelrod1981} has individuals simply play what their opponent played last. This is a relatively high-scoring strategy for individuals to play, although it can be defeated \citep{Nowak1993}. These results extend the current model by allowing individuals to deceive their opponent regarding their previous play by using a probability of telling the truth ($P_{truth}$). In this spatially-explicit game cells are replaced by relatively high-scoring neighbors while their $P_{truth}$ can evolve over time. The contribution of this model is not only to examine what behavior might be favored (cooperate or defect) but investigates how truth-telling affects the outcome of long-term interactions.

A model by Worden and Levin (\citeyear{WordenLevin2007}) explored the ability of individuals to evolve their strategy of playing C or D and conclude that it is possible to escape the prisoner's dilemma when individuals evolve their strategies. This model does not allow for this but instead explores what happens when populations evolve the ability to either lie or tell the truth. This seems more likely that individuals can learn, for instance, to lie when it becomes beneficial.

When individuals simply play their default strategies the advantages of defecting become apparent. This is seen when communities are populated with an even distribution of C and D individuals (figure \ref{grid.fig}). Cooperators in those communities are rapidly replaced by defectors. If instead, individuals play the tit-for-tat strategy then they query their opponents to find out what they played last and then play that themselves. The results depend strongly on the $P_{truth}$ parameter allowing, for instance, communities of lying cooperators to recover to pure truth-telling, cooperative communities. 

Perhaps most intriguing is how scores increase for both cooperators and defectors as truth-telling evolves toward telling the truth or lying, respectively (figure \ref{score-vs-truth}). This means that communities of either truth-telling cooperators or lying defectors are stable. To test these conditions, different individuals were introduced into these communities and found that invasion is not possible. However, not all community types are uninvadible (figure \ref{invaders.fig}). Perhaps most interesting is that communities of truth-telling defectors ($D_T$) and lying cooperators ($C_L$) can both be invaded by a single truth-telling cooperator (figure \ref{invaders.fig}).

In addition, the community filled with lying defectors cannot be invaded by any other strategy (figure \ref{invaders.fig}). The question then leads to what happens when a community is made up of a random assemblage of C and D individuals that are playing TFT and the truth-telling parameter can evolve over time? There exists a threshold at an initial truth-telling parameter of 0.75, and above, where the community will increase to truth-telling cooperators and below which the community decays into a population of lying defectors.

Can we contemplate scenarios in human behavior, for instance, in which these results, based on a relatively simple model, might apply? In political systems where telling the truth is valued we can see that cooperation should be favored. The US Constitution (https://www.senate.gov/about/origins-foundations/senate-and-constitution/constitution.htm), for instance, states in its preamble that its purpose is to ``form a more perfect Union, establish justice, insure domestic tranquility, provide for the common defense, [and] promote the general welfare [of its people].'' This clearly  describes a population of cooperators that are assumed to tell the truth ($C_T$). However, this model suggests that this system should not be stable since one, or perhaps just a few, defecting liars ($D_L$) can invade such a republic and reduce it to a system of lying defectors (figure \ref{invaders.fig}). 

In contrast, is it possible for a community filled with defectors to become a community of cooperators? Unfortunately, this model does not assess the emergence of cooperators when none exist (strategies do not mutate). However, if the population contains a mix of strategies then, if truth telling remains relatively high ($P_{truth} \geq 0.75$), the population will move toward truth-telling cooperators. So, if global cooperation is desirable, then we might hope that lying is not more widespread than 25\% of the population.

Another consideration is that this model could represent the dynamics between species that cooperate and, possibly, signal their intention to cooperate but may, in fact, deceive their partners. Such an interaction exists between cleaner fish and their ``clients.'' Both the cleaner fish and the client generally signal their intentions to cooperate (the client arrives at a cleaning station). However, the cleaner can, instead of cleaning off parasites, defect and ingest part of the fish. Additionally, the client could defect by consuming the cleaner \citep{Poulin1996, Gingins2013}. Interestingly, clients generally truthfully signal their intentions to be cleaned and their readiness to depart the cleaning station through particular body movements \citep[][as cited in Trivers, \citeyear{Trivers1971}]{EiblEibesfeldt1955}. This allows the cleaner fish to exit and move to a safe location.

An apparent cooperative interaction has recently been documented between Pacific white-sided dolphins (\textit{Lagenorhynchus obliquidens}) and fish-eating killer whales (\textit{Orcinus orca}) \citep{Fortune2025}. The dolphins apparently help the killer whales locate schools of large Chinook salmon (\textit{Oncorhynchus tshawytscha}) that are preyed upon by the whales. The dolphins then feed on the remains of the salmon killed during the hunt. Killer whales are known, however, to prey on dolphins, but this apparently does not occur in this pod. Such defection by the orcas would likely lead to an end in the cooperative relationship.

Another, more ancient predator-prey relationship that now functions mutualistically is that of eukaryotic organisms. Early in the evolution of single-celled organisms the prey of one type of organism was likely eventually retained unharmed and would then, instead, function as mutualistic organelles, a phenomenon referred to as endosymbiosis \citep{Sagan1967}.

A final example includes the evolution of a cooperative behavior observed within the bacterium (\textit{Pseudomonas aeruginosa}) \citep{Griffin2004}. Populations of this bacterium tend toward cooperative behavior (producing iron-binding siderophore compounds) when interacting with highly related populations as opposed to more unrelated populations. The cooperative behavior breaks down, however, when individuals are interacting locally and, therefore, competing for scarce resources.

%--------------------------------------------------------------------
\section{Acknowledgements}

I would like to thank the developers of the R Language and Environment for Statistical Computing \citep{R} and the Posit Team for maintaining RStudio (\citeyear{RStudio}).

\end{document}